# A Multiplication-Free Feature Extractor for Signal Classification: Keyword Spotting Case Study

Radu Dogaru, *Member, IEEE*, Ioana Dogaru

***Abstract*— A very low complexity feature extractor called next iRDT is proposed and evaluated for the problem of keyword spotting (KWS). Unlike any other types of feature extractors including the widely used MFCC, or adaptive, CNN-based ones, our algorithm is multiplier-free and it employs only simple, energy-efficient arithmetic operators. Since keyword-spotting of speech commands (KWS) is a typical application for TinyML platforms requiring low complexity for the signal classification chain, we consider it as a case study to evaluate complexity and functional performance. If properly tuned, iRDT demonstrates similar accuracy to solutions based on MFCC or CNN-based extractors using baseline classifiers on Google's KWS 12-classes dataset. With a different classifier, the system achieved 94.7% validation accuracy. Processing times on CPU for the proposed feature extractor, are at least one order of magnitude smaller than for the MFCC. The proposed algorithm has a very low hardware footprint, making it ideal for ultra-low-power edge devices. Code and demo are available [18].**

## I. Introduction

FEATURE EXTRACTION is an essential step in a signal classification chain. In most cases, extractors such as Mel-Frequency Cepstral Coefficients (MFCC) [1], or Mel-Filter bank Energies (MFE) producing 2-dimensional spectrograms are combined with convolutional neural network (CNN) classifiers for maximal accuracy. Specific applications such as Keyword Spotting [2] are associated with portable assistants and require deployment on TinyML computation platforms [3-5] where energy consumption and latency are issues of crucial interest. Specialized design frameworks for TinyML platforms such as EdgeImpulse [6] offer support only for spectral-type feature extractors (e.g. MFCC) and it is widely accepted among TinyML developers that in a signal classification flow, the most computation takes place in the feature extractors. For instance, in [7] authors report 300 ms latency for the MFCC and only 2ms for the classifier. The computational complexity of MFCC is associated with multipliers and cosine operators inside the fast Fourier transform (FFT). Recently, in [8] authors propose an automatic feature extractor (FE) which is essentially a CNN autoencoder trained with signals from Google's KWS dataset [11]. The 2-dimensional FE is extracted from a hidden layer of the network and according to authors, compares favorably with MFCC and other spectral FEs. Still, the proposed FE needs intensive training, relies on rather complex convolutional operators (including multiplication) and must be retrained when applied to other classes of signals.

Both authors are with Dept. of Applied Electronics and Information Engineering, National University of Science and Technology POLITEHNICA Bucharest, Romania (e-mail: corresponding author radu.dogaru@upb.ro , ioana.dogaru@upb.ro); First author is also Associate with Technical Sciences Academy of Romania.

In this letter, building upon our previous work [9], we propose a multiplier-free alternative to traditional 2D-spectrograms used in signal classification and demonstrate it performs equally well, in terms of accuracy, to more sophisticated feature extractors (including MFCC and FE in [8]) for the case study problem of keyword spotting. For fair reference, the same setup (dataset and classifier) as in [8] is considered, to have a basis for comparison. Section II presents the algorithm and its hyperparameters as well as rules for tuning them to achieve best balance between complexity and accuracy. Section III details on employing the proposed FE, called next iRDT, in signal classification chains where similar accuracies to ones reported in [8] are obtained. Using other small foot-print classifiers [10] improves validation accuracies up to 94.7% for the 12 classes Google's KWS dataset [11].

Compared to the computationally heavy MFCC, our approach has a minimal hardware/energy footprint [12] and when hardware-oriented platforms are considered it would require very few logic gates (LUTs/FFs in an FPGA). Its simple description and the lack of multipliers ensure facile implementation on various computational platforms, enabling convenient signal classification applications on low energy and cost platforms.

## II. The Proposed Feature Extractor

In [13] we first propose expanding a 1D-Laplacian based descriptor (inspired by reaction-diffusion cellular nonlinear networks, previously used to classify emergent dynamics in cellular automata [14]) to signal analysis, under the name of RDT (reaction-diffusion transform). Initially it was applied as a 1D-feature extractor in conjunction with support vector classifiers and proved quite efficient for speech classification problems. Later [9] the concept was expanded into a 2D, spectrogram-like, feature extractor and demonstrated comparable performance in various signal classification problems including environmental and emotional speech recognition [15], or EEG classification in TinyML [16]. The version considered herein is a heavily optimized and improved version of the one proposed in [9] (almost 60 times faster), henceforth called improved RDT (abbreviated next iRDT). A comparison with the widely used MFCC feature extractor is given in Figure 1. Specific to iRDT is the use of very simple operators, the basic computational unit including only addition/subtraction, absolute value and shift (multiplication by 2) specific to the 1D Laplacian. The feature size in both cases is *(Mxm)* where *M* represents the integer number of frames (segments) splitting the entire sequence $s(t), t = 1,..N$ and *m* is the number of cepstral coefficients for MFCC, or the number of delays $d_k, k = 1,..m$ for iRDT.

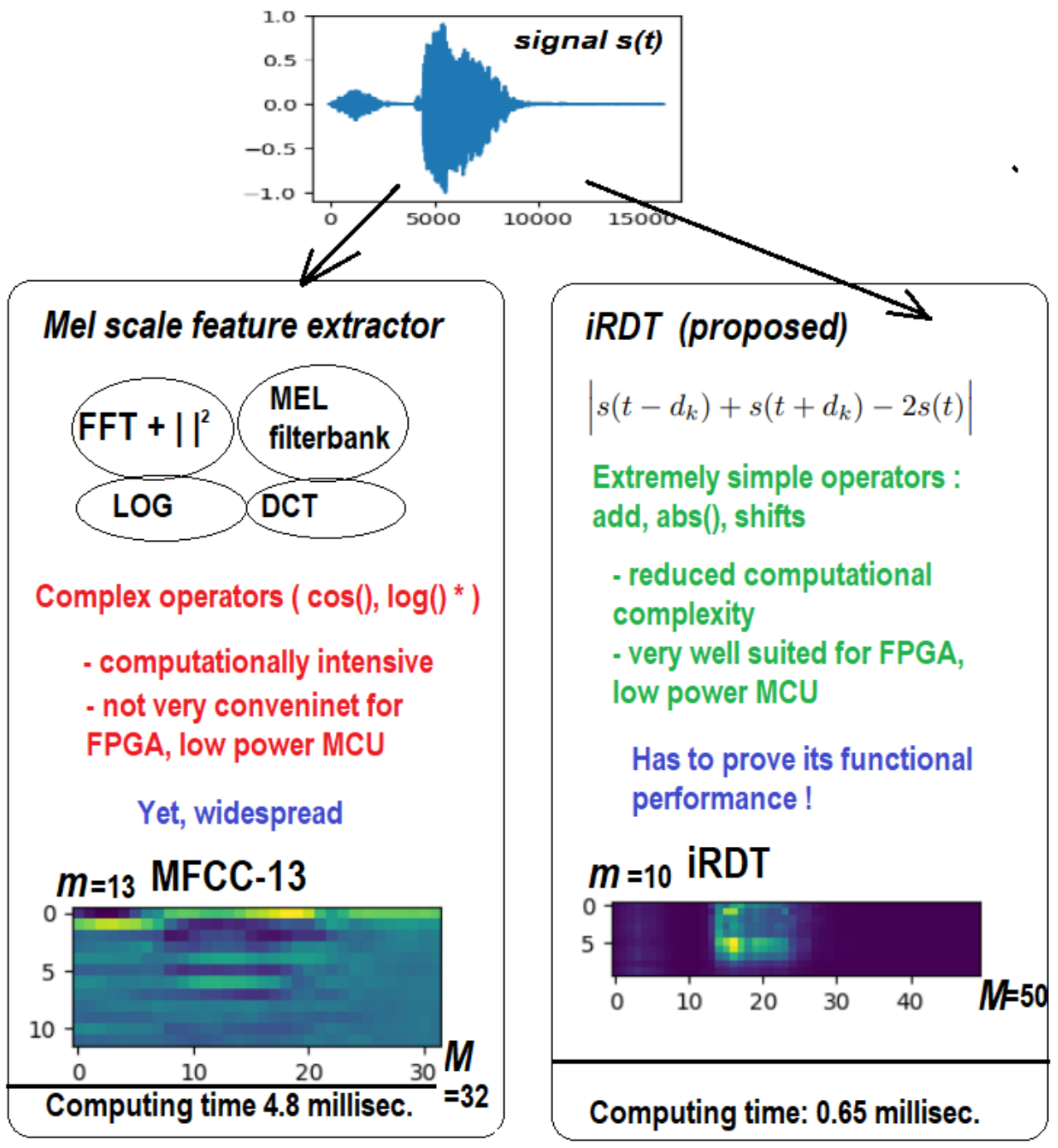

Figure 1. A comparison between traditional (MFCC) and iRDT feature extractors.

Due to the very simple operators involved in iRDT, similar sizes of features are computed much faster than with MFCC.

### *A. Algorithm and its hyperparameters*

There are two versions of the iRDT algorithm: i) the first, called **iRDTf** (f stands from fixed *M*) prescribes *M* independently of the signal length *N*. It is convenient in cases where signals associated with different classes are variable. In the case of KWS we consider the dataset in [17] where sounds are in general fixed size corresponding to 1 second sequences sampled at 16 Khz (i.e. *N=16*). Consequently, in this work we emply the second version: ii) **iRDTv** – (v stands from variable *M*) i.e. the number of frames is determined by the length *N* of the signal. The algorithm is listed next and it corresponds to the Python function **irdtv()** available in the associated code and demo [18]. Our implementation is optimized using JIT (just in time) compiler from NUMBA library, as well as the MFCC from LIBROSA library, considered as MFCC reference in our code.

**Algorithm 1** iRDTv (variable number $M$ of segments)

**Require:** signal, chan, win_per_segm, $w$
**Ensure:** $M$, Feat_spec

1: $m \leftarrow$ length(chan), windows $\leftarrow \lfloor$length(signal)$/w\rfloor$
2: $M \leftarrow \lfloor$windows/win_per_segm$\rfloor$
3: Feat_spec $\leftarrow$ zeros$(M, m)$ ▷ 32-bit int
4: spectrum $\leftarrow$ empty(windows, $m$) ▷ 32-bit int
5: limit $\leftarrow \lfloor w/4\rfloor$, base $\leftarrow 0$ ▷ In MCU implementations lines 1-2 (divisions) are precomputed

6: **for** $i = 0$ to windows $- 1$ **do**
7: **for** $k = 0$ to $m - 1$ **do**
8: delay $\leftarrow$ chan$[k]$, spik $\leftarrow 0$
9: **for** $t =$ limit to $w -$ limit $- 1$ **do**
10: spik $\leftarrow$ spik $+$ |signal[base $+ t -$ delay] $+$ signal[base $+ t +$ delay] $-$ 2signal[base $+ t$]|
11: **end for**
12: spectrum$[i, k] \leftarrow$ spik
13: **end for**
14: base $\leftarrow$ base $+ w$
15: **end for**

16: base $\leftarrow 1$ ▷ To avoid padding, ignore first/last spectral lines
17: **for** $k = 0$ to $M - 2$ **do**
18: **for** col $= 0$ to $m - 1$ **do**
19: F_k_col $\leftarrow 0$
20: **for** $i =$ base to base $+$ win_per_segm $- 1$ **do**
21: F_k_col $\leftarrow$ F_k_col $+$ spectrum$[i,$ col$]$
22: **end for**
23: Feat_spec$[k,$ col$] \leftarrow$ F_k_col
24: **end for**
25: base $\leftarrow$ base $+$ win_per_segm
26: **end for**

27: **return** $M$, Feat_spec

The inputs of the algorithm are the signal sequence *s(t),* a list of integer delays $chan = [d_1, d_2, \; .. \; d_k, .. d_m]$ the sliding window size *w* and an integer number *win_per_segm* representing the number of sliding windows per frame (segment). The algorithm outputs the number of frames *M* and an array of size *Mxm* containing the 2D-feature. The main computation takes place in line 10 and consists of 9 simple operations (4 additions, one shift for the multiplication by 2, an absolute value, and 3 loads from SRAM). The loop starting with line 7 and ending with line 13 computes for each sliding window *i* a "spectral" line *spectrum* (the RD transform of that window) with *m* channels. For the choice *lim=w/4* there are $9m\left(\frac{w}{2}\right) = 4.5mw$ basic operations per loop. At the end of main loop (line 15) all spectrums for all *N/w* sliding windows are computed giving an upper estimate for the overall number of elementary operations of the iRDTv:

$$Ops(iRDTv) \cong 5mN \qquad (1)$$

The remaining lines (loop 17-26) would average spectrums per each of the resulting *M* frames to finalize the computation of the spectrogram. The number of operations involved here is negligible with respect to computations in the main loop (lines 7-13), consequently we may conclude that equation (1) gives a good estimate for the number of elementary operations of the iRDT transform. For instance, assuming *N=16000*, *m=6* (a typical setup in the following experiments) it follows that iRDT would have a complexity of 480 kilo-Flops (in fact integer int32 ops). A rough estimate of MFCC according to procedure in [19] gives (for the case considered in this work, with FFT window size=2048, Hop size = 512, number of Mel coefficients m=13) around 3 Mega-flops (including multiplications and harmonic functions, taking more clock cycles per operation) for the same signal.

### B. Classification chain and datasets

The classification chain is depicted in Fig.2.

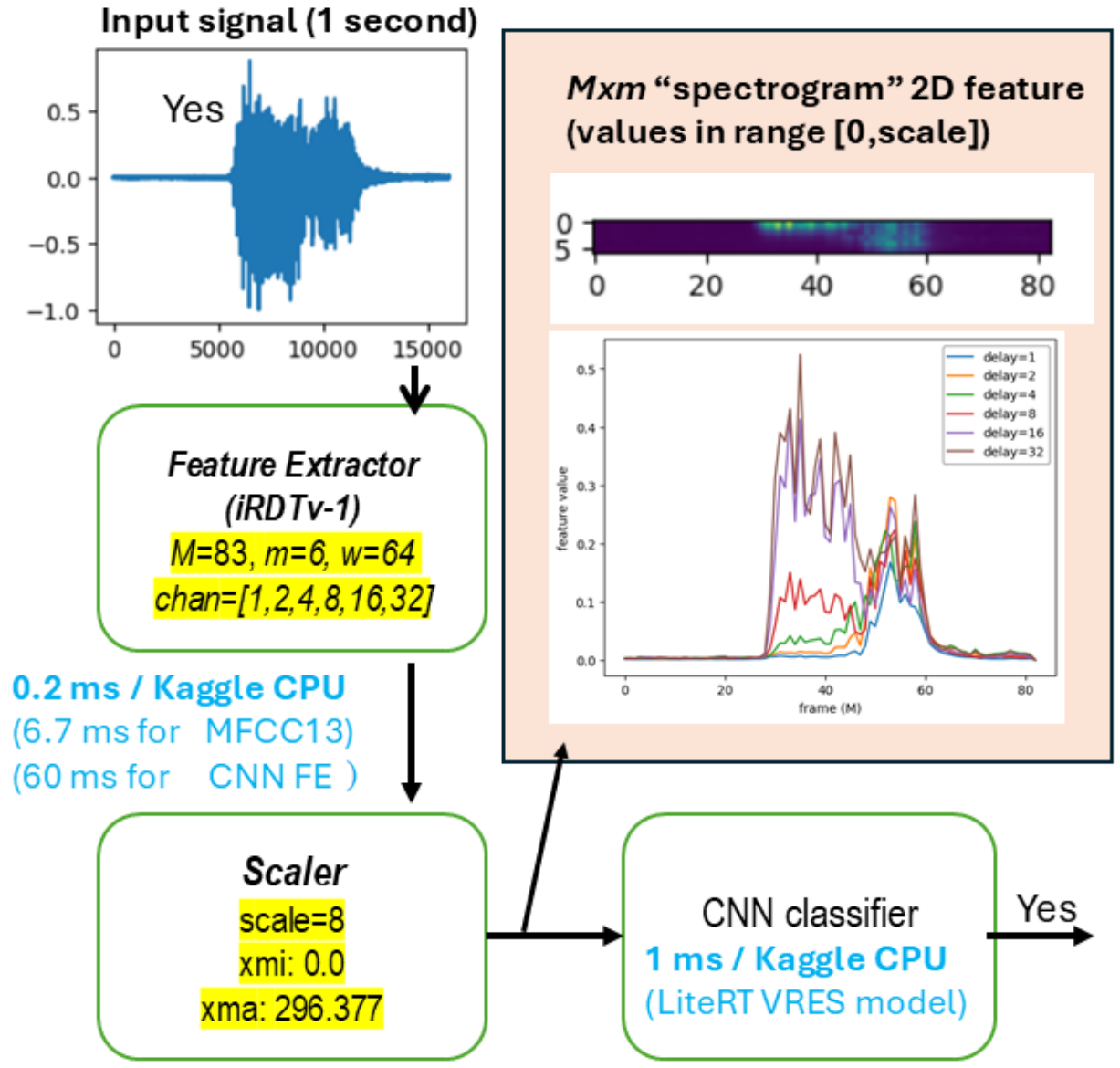

Figure 2: Signal classification chain.

As seen, the chain includes the iRDT feature extractor (or other feature extractors (FE), for comparisons) and a scaler providing the input of the classifier with normalized and scaled data. The values *xmi* and *xma* are produced by the scaler after providing all signal sequences in the training set to the FE and they should be stored and used in the retrieval phase. The dataset used was prepared similarly to the one described in [8] with 80% of the samples used for training and 20% for validation. The entire set is based on [17] where in addition to 10 speech commands, 2 classes with the same number of signal samples per class (3772) were added: "silent" representing 1-second-long background noise sequences selected randomly from the original set and "unknown" with an equally-balanced mix from the remaining commands in the original set. The 12 classes were labeled as follows: 0:Yes, 1:No, 2:Down, 3:Up, 4:Left, 5:Right, 6:Off, 7:On, 8:Go, 9:Stop, 10:Silent, 11:Unknown. For faster tuning of the FE hyperparameters a reduced version of the set with only 200 samples per class was employed. Two classifiers were considered in our study: The first, is the DS_CNN [20] (using the implementation from [21]) considered as reference in baseline paper [8]. The second, entitled VRES-CNN, is available from [22] with a detailed description in [10]. The structure of VRES used herein is with *flat*=0, *fil*=[40,100,60,30], *nl*=[2,1,1,0], *hid*=[], and convolutional kernels sized *5x3* instead of *3x3* as in the original work.

### C. Tuning iRDT for best signal classification performance

There are at least 3 hyperparameters to tune in iRDT: **i) the choice of *w*** – in practice *w* is chosen as a power of 2 (although this is not compulsory) and quite usually its choice is such that in the end the number of frames is large enough to get good accuracy. For the KWS problem, our selection was *w*=64; **ii) the delay list *chan*** is a list of powers of 2 ending with $d_m \leq w/2$, consequently default chan=[1,2,4,8,16,32] and *m=6.* Intermediate values may be also inserted in the list of delays, as in some cases they may improve the classification accuracy at the expense of increasing complexity; **iii) the most critical parameter is *win_per_seg*** who directly influences the number of frames *M* which in effect, would influence the classifier complexity. A seen in Table 1, a large *M* value generally leads to better accuracy.

TABLE I – OPTIMIZING IRDT FEATURE EXTRACTOR FOR BETTER OVERALL CLASSIFICATION ACCURACY

| iRDT version | Win_per_segm | M | m | Feature size *Mxm* | Acc(%) DS_CNN | Acc(%) VRES |
|---|---|---|---|---|---|---|
| A0 | 2 | 125 | 6 | 750 | 89.79 | 91.87 |
| A1 | 3 | 83 | 6 | 498 | 89.37 | 91.45 |
| A2 | 4 | 62 | 6 | 372 | 86.87 | 91.04 |
| A3 | 5 | 50 | 6 | 300 | 87.29 | 88.54 |
| A4 | 6 | 41 | 6 | 246 | 87.29 | 90 |
| B1 | 3 | 83 | 9 | **747** | **91.87** | **92.29** |
| B2 | 4 | 62 | 9 | **558** | **91.45** | 90.62 |
| B3 | 5 | 50 | 9 | 450 | 88.75 | 91.04 |
| B4 | 6 | 41 | 9 | 369 | 88.12 | 90.2 |
| B5 | 7 | 35 | 9 | 315 | 87.5 | 89.58 |
| C1 | 2 | 62 | 7 | 434 | 86.45 | 89.16 |
| C2 | 3 | 41 | 7 | 287 | 85.83 | 89.58 |

In the above table, 3 types of *chan* lists were used: A) the default list for *w*=64 i.e. *chan*=[1,2,4,8,16,32]; B) some additional delays inserted: *chan* = [1,2,4,6,8,12,16,24,32] and C) the default *chan* list for *w*=128. For each type, several versions denoted by figures were considered, each being associated with a specific *win_per_segm* choice implying various FE sizes. The validation accuracies obtained when using both DS_CNN and VRES classifiers on the reduced dataset after 200 epochs (batch size = 48) are listed in the last two columns.

Note that for a given feature size (the smaller the better in terms of complexity) FE models with B type of channels provide best performance, while the choice of *w*=128 gives relatively low performance. Lower complexity is ensured with type A channels. In the next section the models emphasized in Table 1 with colored background (denoted next as iRDTv-A2, B1, B2, B3) would be further investigated for the complete dataset.

## III. Experimental Study and Results

### A. Comparison with baseline research

Table II summarizes the results of using various iRDTv FE versions (denoted as B1, B2, B3 and A2) along with the Autoencoder-FE and MFCC in baseline work [8] and our MFCC FE (implemented via LIBROSA library, as mentioned previously). From left to right the FE are ordered in decreasing order of the feature size (*Mxm*), and the Kilo-MAC (Multiply and Accumulate) complexity of both classifiers is considered, estimated with the Tensorflow profiler (tf.profiler) tool. Better accuracies are obtained with larger feature sizes at the expense of increasing complexities for both FE and the classifier.

Table II: Overall comparison of signal classifiers using various iRDT versions and other feature extractors (when 8-bit outputs are used in the FE, accuracies are given with *italic* font)

| **Feature extractor** | FE in [8] | iRDT v-B1 | iRDTv-B2 | iRDT v-B3 | MFCC-13 | iRDTv-A2 |
|---|---|---|---|---|---|---|
| **Feature size** | 800 | 747 | 558 | 450 | 416 | 372 |
| **DS-CNN Acc (%)** | 90.36 [8] | 92.11<br>*92.11* | 91.08<br>*91.89* | 91.47<br>*91.77* | 91.7<br>91.04 [8]<br>*91.06* | 91.05<br>*91.47* |
| **VRES Acc (%)** | - | 94.73<br>*94.47* | 93.53<br>*94.07* | 93.27<br>*93.83* | 94.4<br>*94.13* | 92.25<br>*93.23* |
| **DS_CNN K-MAC** | - | 4651 | 3433 | 2769 | 2482 | 2061 |
| **VRES K-MAC** | - | 7807 | 5812 | 4695 | 4212 | 3694 |

It follows that using iRDT gives comparable accuracy with those reported in baseline [8] where DS-CNN classifier was used. Moreover, when using VRES instead of DS-CNN, accuracies up to 94.7% are achievable, with some increase in complexity. Compared to 91.04% reported in [8] for MFCC, our MFCC gives better accuracy (91.7%) on the same classifier, possibly because 8-bit quantified feature was used there (case when our results are also close i.e. 91.06%). A fair comparison would consider the 8-bit quantified FE discussed in next sub-section.

### B. Improving classification accuracy for 8-bit quantization

Since TinyML usually requires the compressed implementation of classifiers, often using INT8 (8 bit) quantization, we were interested to see how the overall accuracy is influenced when 8-bit quantization is applied to spectrograms generated by the considered FE. It turns out that for the iRDT the accuracies drops are in the range 2 to 2.5% while in the case of our MFCC the drops are much smaller (under 0.6%). To alleviate this problem, the logarithmic transform $log_2(1 + Feat_spec)$ is next applied to the output of the iRDT feature extractor, similarly to the logarithmic compression in the MFCC. The effect is positive, now the quantization drops were negligible (under 0.1%) and validation accuracies improved in most cases. The accuracy for 8-bit quantized outputs of the FE is given in italics and blue color in Table II. Adding the logarithm (implemented in MCU platforms quite efficiently with precalculated lookup tables stored in Flash memory) gives no major increase in the overall computational complexity. Even in the case of using the most aggressive log2 approximation (bit-shifting, giving spectrograms composed only by integers in range 0 to 9 at the output, the accuracy drop is less than 3%). In terms of 8-bit quantified FE, the iRDT + DS-CNN classifier performs better than MFCC+DS-CNN even for smaller feature size (91.47% with FE size 372 versus 91.06% with FE size of 416) and allows getting up to 94.47% accuracy in setups with higher computational complexity.

The Python code of the proposed FE is available [18] and a demo running on Google Collaboratory allows testing the functionality based on several pretrained classifier .tflite models. Figure 3 gives a comparison of 3 FE choices, and it is interesting to note execution times for iRDT (0.2 – 0.3 milli-seconds) versus MFCC with 6-7 milliseconds.

**iRDTv1: w=64, chan=[1,2,4,8,16,32], M=3 (83x6)**

```
Real class :  2Down
Predicted class:  1
False
iRDTv1 feature extractor latency:  0.2089 milli-seconds
classifier latency:  1.04 milli-seconds
100 experiments:
6 in 100 failed: test_accuracy= 94.0 %
```

**iRDTvopt: w=64, chan = [1, 2, 4, 6, 8, 12, 16, 24, 32] M=3, feature size: (83x9)**

```
Real class :  1No
Predicted class:  1
Correct
iRDTvopt feature extractor latency:  0.3153 milli-seconds
classifier latency:  1.37 milli-seconds
100 experiments:
1 in 100 failed: test_accuracy= 99.0 %
```

**MFCC13 : mels=13, FFT=2048, HOP size=512**

```
Real class :  3Up
Predicted class:  3
Correct
MFCC13 feature extractor latency:  6.4111 milli-seconds
classifier latency:  1.51 milli-seconds
100 experiments:
3 in 100 failed: test_accuracy= 97.0 %
```

Figure 3: Three FE experiments with the demo code.

The user can select one of three available FE (including pretrained classifiers) and enter a loop running more experiments. Each experiment selects randomly a signal sample and follows the steps indicated in Fig.2 showing processing times, decision and some estimate of the accuracy given for the samples selected in this "ad-hoc" test set. The user can also test the functionality with novel signals (out of the KWS set).

## IV. Conclusion

A novel, multiplier-less algorithm for feature extraction with relevance in signal classification, was introduced and evaluated for the case study of keyword spotting. It can act as a replacement for the widely used MFCC or more recent CNN-based FE embeddings, without compromising overall classification accuracy but gaining an important reduction in complexity for both computing time and hardware resources. The very simple algorithm can be easily integrated into a wide variety of platforms, and it has the possibility for best tuning to achieve a good balance between accuracy and complexity of the overall system. Further research would investigate its use for other signal classification problems.